\documentclass[smallextended]{svjour3}       

\usepackage{amsmath}
\usepackage{amsfonts}
\usepackage{amssymb}
\usepackage[square,comma,numbers, sort & compress]{natbib}
\usepackage{mathrsfs}
\usepackage{graphicx}

\begin{document}

\title{Inflation and reheating in scale-invariant scalar-tensor gravity.}
\author{Giovanni Tambalo \and Massimiliano Rinaldi}

\institute{G. Tambalo \at
              SISSA \\
              via Bonomea 265,\\
              34136 Trieste, Italy\\
              \email{gtambalo@sissa.it}           
           \and
           M. Rinaldi \at Physics Department \\ University of Trento \\ Via Sommarive 14, \\38123 Povo (TN), Italy \and
              TIFPA - INFN \\ Via Sommarive 14, \\38123 Povo (TN), Italy\\
\email{massimiliano.rinaldi@unitn.it}   
}

\date{Received: date / Accepted: date}

\maketitle

\begin{abstract}
We consider the scale-invariant inflationary model studied in \cite{rinaldi_inflation_2016}. The Lagrangian includes all the scale-invariant operators that can be built with combinations of $R, R^{2}$ and one scalar field. The equations of motion show that the symmetry is  spontaneously broken after an arbitrarily long inflationary period and a fundamental mass scale is  generated. Upon symmetry breaking, and in the Jordan frame, both Hubble function and the scalar field undergo damped oscillations that can eventually amplify Standard Model fields and reheat the Universe. In the present work, we study in detail inflation and the reheating mechanism of this model in the Einstein frame and we compare some of the results with the latest observational data.

\keywords{Inflation \and Modified gravity \and Reheating}
 \PACS{98.80.Cq \and 04.50.Kd \and 04.62.+v}
\end{abstract}

\section{Introduction}

The presence of an inflationary epoch \cite{guth_inflationary_1981} in the early history of our Universe is widely considered as a necessary requirement for any realistic cosmological model. This has been supported by several observations \cite{tonry2003cosmological,riess2007new,komatsu2011seven} that can be explained only if the metric undergoes a stage of exponential expansion for several e-folds. Recently, we entered in a phase of high-precision cosmological measurements, such as the ones of Planck \cite{planck_collaboration_planck_2015}, which put strong constraints on inflationary models, and are able to rule out many competing proposals.

Scale-invariant models of gravity are a source of inspiration for many inflationary mechanisms since they are  able to predict for the spectral index of scalar perturbations $n_s$ values close to 1 and thus consistent with current observations. However, in order to fit with all observables, such symmetry cannot be exact. In other words, symmetry must be broken dynamically or by the introduction of small non-invariant terms in the action, usually justified by quantum corrections (for example see \cite{zee1979broken,starobinsky_new_1980,cooper1981cosmology, gorbunov2014scale,turchetti1981gravitation,Salvio2014,rinaldi2015inflation,rinaldi2016quasiscale,barrie2016natural}).
For other scale invariant cosmological models see \cite{padilla2014generalized,saridakis2016bi,Kannike2015,Kannike2016}.

A recent proposal, where scale-invariance appears as a global symmetry, is the classical scalar-tensor theory studied in \cite{rinaldi_inflation_2016}. Here, the Lagrangian is composed by scale-invariant operators built on combinations of $R$, $R^{2}$, and a  scalar field $\phi$ with its standard kinetic term. The dynamical analysis of the equations of motion in the Jordan frame reveals that the system has only two fixed points. The first, unstable, correspond to a (quasi) de Sitter spacetime with an arbitrary small scalar field. The second, stable, correspond to damped oscillations of the Hubble parameter and of the scalar field around fixed values. The path from the unstable to the stable point corresponds to an arbitrarily long inflationary phase (depending loosely upon the initial conditions) followed by  the damped oscillations. The equilibrium value of $\phi$ determines  a fundamental mass scale, which emerges dynamically. Thus the breaking of the global scaling symmetries is able to generate a mass scale that can be identified with the Planck mass. Finally, the oscillations of the Hubble parameter and of the scalar field finally allow the excitation of the Standard Model fields and the reheating of the Universe. The results obtained in \cite{rinaldi_inflation_2016} show that observables are consistent with observations, at least in the Jordan frame.

In this paper we aim at analyzing carefully the dynamics of the system in the Einstein frame, where the comparison with more conventional inflationary models is straightforward. In particular, we focus on two aspects. The first is the inflationary phase: in the Einstein frame formulation we have two scalar fields at play (one is the so-called scalaron, springing from the metric redefinition) and, technically speaking, the model belongs to the class of hybrid inflation \cite{linde1994hybrid}.

However, from scale invariance we desume that one combination of the two scalar fields is to be considered, at the classical level, as a ``spectator'' field, so it is the remaining field only that drives the quasi-exponential expansion. With this observation, we compute the spectral indices and compare them to Planck data. The second aspect concerns preheating. In the Einstein frame, both the scalaron and the original scalar field undergo damped oscillations at the end of inflation. In the hypothesis that the Standard Model fields are coupled to these oscillating quantities, we show that there are at least three different and efficient particle production channels that can reheat the Universe after inflation. 

This paper is organised as follows. In section \ref{II} we present the main features of the model studied  in \cite{rinaldi_inflation_2016} together with the formulation in the Einstein frame (which results useful in the calculation of cosmological parameters). In \ref{III} we study, through a fixed point analysis of the equations of motion, the dynamical evolution of the model in a FLRW metric. Subsequently, in section \ref{IV} the inflationary analysis is carried out at second order in the slow-roll parameters. This allows the comparison of our observables with the Planck data. In section \ref{V} we show how the model allows for an efficient energy transfer from inflationary fields to matter fields in the so-called preheating scenario. We conclude in \ref{VI} with some considerations.

\section{Scale-invariant inflationary model}\label{II}

The action considered in \cite{rinaldi_inflation_2016} reads
\begin{equation}\label{model_action}
I = \int \left[\frac{\alpha}{36}R^2 + \frac{\xi}{6}\phi^2 R - \frac{1}{2} \, \partial_\mu \phi \partial^\mu \phi -\frac{\lambda}{4}\phi^4  \right] \sqrt{-g}\, \text{d}^4x\,,
\end{equation}
with $\alpha$, $\xi$, and $\lambda$ being positive dimensionless parameters, and $\phi(x)$ a real scalar field.  The action \eqref{model_action} is invariant under the active dilatation transformation
\begin{equation}
x' = \ell^{-1} x, \hspace{0.4cm} g'_{\mu \nu} (x) = g_{\mu \nu} (\ell x ), \hspace{0.4cm} \phi'(x) = \ell \phi(\ell x )\,,
\end{equation}
where $\ell > 0 $. Additionally, the rigid Weyl transformation 
\begin{equation}\label{eq:weyl}
 g'_{\mu \nu} (x) = L^2 g_{\mu \nu} (x), \hspace{0.4cm} \phi'(x) = L^{-1} \phi(x)\,,
\end{equation}
with $L > 0$ also leaves $I$ unchanged. The case $L < 0$, in principle admissible, results to be nothing but a combination of a Weyl and a $\mathbb{Z}_2$ transformation $\phi\rightarrow-\phi$. 

Finally, the combined symmetry transformation parametrized by $(\ell, L)$ spans a two-dimensional Abelian group. Evidently, the two underlining symmetries are related by the invariance of the action under coordinate transformation, realised by a combined transformation with $\ell = L$. Thus, a mechanism able to break the Weyl symmetry (but leaving the diffeomorphism invariance unaffected) will inevitably break the dilation symmetry and vice versa. 
As a consequence, the action cannot contain a cosmological constant or any other coefficient with dimensionality different from zero.

The classical effective  potential for the scalar field corresponds to
\begin{equation}
\mathscr{V} (\phi)=\frac{\xi}{6}\phi^{2}R-\frac{\lambda}{ 4}\phi^{4}\,.
\end{equation}
When $R$ is constant, it has one local maximum and one local minimum, respectively located at
\begin{equation}
\phi = 0, \hspace{0.5cm} \phi_0 = \pm \sqrt{\frac{\xi R}{3 \lambda}}\,.
\end{equation}
This structure guarantees the presence of a symmetry breaking mechanism in the model. Indeed, it has been  shown  in \cite{rinaldi_inflation_2016} that, in a Universe with infinite spacetime volume and constant $R$, the scale symmetry is broken whenever the field $\phi$ relax to one of the minima and generates a mass scale identified by $\phi_0$. In particular, it has been assessed that this instance occurs in a flat Friedmann-Lemaitre-Robertson-Walker (FLRW). 

A necessary requirement for \eqref{model_action} to describe post-inflationary physics is to reduce to the standard Einstein-Hilbert action after symmetry breaking. Thus, the quartic  self-interaction term for the scalar field and the quadratic term for the Ricci scalar need to cancel out, implying $\alpha = \xi^2 / \lambda$. The model, therefore, is left with only two free parameters. Finally, the non-minimal coupling term in the action, at the stable point, reduces to $M_p^2 R / 2$ (where $M_p$ is the Planck mass) provided that $M_p = \sqrt{\xi / 3} \phi_0$.

As any other scalar-tensor theory, the model under consideration can be brought to the Einstein frame through a redefinition of the metric of the form $g_{\mu \nu}(x) \rightarrow \Omega^2(x) g_{\mu \nu}(x)$ for some well-behaved function $\Omega(x)$. In this frame the inflationary analysis can be performed with standard techniques \cite{futamase_chaotic_1989,de_felice_fr_2010,felice_chaotic_2011}.

To proceed in this direction we write the action \eqref{model_action} in the equivalent  representation
\begin{equation}
\label{action_linear_representation}
I = \int \left[ \eta R - \frac{1}{2} \,  \partial_\mu \phi \partial^\mu \phi - \frac{9}{\alpha}\left( \eta - \frac{\xi}{6}\phi^2 \right)^2  - \frac{\lambda}{4}\phi^4 \right] \sqrt{-g}\, \text{d}^4x
\end{equation}
where $\eta$ is a  new scalar field  with mass dimension $2$. It should be clear that $\eta$ is an auxiliary field in \eqref{action_linear_representation} which, along with the equation of motion for $\eta$, is equivalent to the original action \eqref{model_action}. The Einstein frame is obtained with the choice $\Omega^2(x) =  M^{2} / (2 \eta (x))$,  where the arbitrary  mass scale $M$ is introduced for dimensional consistency only \cite{maeda_towards_1989,wands_extended_1994}. As one would expect from the scale symmetry of the model, it can be shown that such scale is a redundant parameter of the action, thus no observable quantity depends on it. In particular, such parameter ought not to be confused with the dynamically generated mass scale after symmetry breaking (which, in the Jordan frame, is identified with the Planck mass). 

Together with the additional redefinition $\chi = -\sqrt{6} M \log \Omega$ we obtain the Einstein frame action
\begin{equation}
\label{conf_model_action}
I_E = \int  \left[ \frac{M^2}{2 }R - \frac{1}{2}\partial_\mu \chi \partial^\mu \chi  - \frac{1}{2} e^{-\gamma \chi}\partial_\mu \phi \partial^\mu \phi - U(\phi, \chi)  - \Lambda M^2 \right] \sqrt{- g} \, \text{d}^4 x\,,
\end{equation}
where the potential $ U(\phi, \chi)$ is defined as
\begin{equation}
\label{pot_einstein_frame}
U(\phi, \chi) = \phi^4 e^{-2 \gamma  \chi } \left( \frac{\lambda}{4} + \frac{\xi^2}{4 \alpha}\right) - \frac{3\xi}{2 \alpha } M^2 \phi^2 e^{-\gamma \chi}\,,
\end{equation}
and where
\begin{equation}\label{cosm_cost}
\Lambda \equiv \frac{9 M^2}{4\alpha }, \hspace{0.5cm} \gamma \equiv \frac{1}{ M} \sqrt{\frac{2}{3}}\,.
\end{equation}
The Einstein frame action has the advantage to disentangle the spin-2 and scalar degrees of freedom of the gravitational sector, whose appearance is explained by the presence of a quadratic term $R^2$ in the Jordan frame action \cite{de_felice_fr_2010}. The formulation of the theory in this frame has the disadvantage of presenting the kinetic term of the scalar field $\phi$ in a non-canonical form, hence the contribution of such field to the total energy is not manifest. 

By varying the action \eqref{conf_model_action} with respect to the conformal metric (indicated by $g_{\mu \nu}$) one obtains the Einstein's equations
\begin{align}\label{conf_model_eq}
&R_{\mu \nu} - \frac{g_{\mu \nu}}{2} R  + g_{\mu \nu} \Lambda = \\\nonumber&= {1\over M^{2}} \left[ \partial_\mu \chi \partial_\nu \chi   - g_{\mu \nu}\frac{(\partial \chi)^2}{2}  + \left( \partial_\mu \phi \partial_\nu \phi - g_{\mu \nu}\frac{(\partial \phi)^2}{2} \right)e^{-\gamma \chi} - g_{\mu \nu}  U \right]\,.
\end{align}
The variation with respect to the scalar fields yields the Klein-Gordon equations 
\begin{align}
\label{eq1_field}
\square \chi& = \frac{\partial U}{\partial \chi}  -  \frac{\gamma}{2} e^{-\gamma \chi} \left( \partial \phi \right)^2\,, \\
\label{eq2_field}
\square \phi& =  \frac{\partial U}{\partial \phi} e^{\gamma \chi} + \gamma  \partial_\mu \chi \partial^\mu \phi \,.
\end{align}
Although these two last equations are highly entangled, the overall system of equations is much more manageable than in the Jordan frame. Complications arise in the Einstein frame solely due to the fact that one kinetic term is not canonical, so additional derivative couplings appear.


In general, different redefinitions of the fields do not alter the underlying physics. Thus, it is reasonable to recast the kinetic part of the action \eqref{conf_model_action} for our two scalars in a way that makes this equivalence manifest. In order to do so, we indicate our set of fields as $\phi^a = \{ \phi^1, \phi^2\}$. Of course, in the case of equation \eqref{conf_model_action} this entails $\phi^a = \{ \chi,\phi \}$ (with $a = 1, 2$). Then, it is possible to introduce a metric $\gamma_{ab} (\phi^1, \phi^2)$ (or in short $\gamma_{ab}$) for the field space in such a way that the kinetic term in the Lagrangian for scalars is written as a scalar product in both the field and the spacetime metric
\begin{equation}\label{eq:L_fully_cov}
\mathscr L_E^{kin} = -\frac{\sqrt{-g}}{2} \gamma_{ab} \partial_\mu \phi^a \partial_\nu \phi^b g^{\mu \nu}
\end{equation}
The transformation properties of $\gamma_{ab}$ under a redefinition of fields are manifest.

For the argument that will follow it is convenient to perform a particular field redefinition  $f \equiv M e^{-\gamma \chi /2} $ thus $\phi^a = \{f, \phi\}$. After a trivial change of variables the metric $\gamma_{ab}$ becomes $\gamma_{ab} = \text{diag} \left( 6 M^2 f^{-2}, f^2 M^{-2}\right)$, and the potential $U(f, \phi)$ becomes just a function of $(\phi f)$. In this formulation it is manifest that with a mere redefinition of fields it is not possible to obtain a canonically normalized kinetic term for the scalars (e.g. $\gamma_{ab} = \delta_{ab}$) since the Riemann tensor associated to the field metric is not vanishing. Specifically, the Ricci scalar $R[\gamma_{ab}] = -1/3$ hence the field manifold is hyperbolic.

Although a trivialization of $\gamma_{ab}$ cannot be achieved, it is still possible to find a set of fields for which the potential takes a more manageable form.
The rigid Weyl symmetry can be exploited to this aim.
In the Einstein frame the analogous of \eqref{eq:weyl}, in infinitesimal form, turns out to be
\begin{equation}
\phi'(x)  = \phi(x)  + \sigma \phi(x), \hspace{0.4cm} f'(x)  = f(x) -\sigma f(x)
\end{equation}
where $\sigma >0 $ represents a small parameter. The metric $g_{\mu \nu}$ in Einstein frame results neutral under the corresponding Abelian group (contrary to what happens in Jordan frame).
Evidently, it is always possible to redefine the fields ${\phi^a}' = \{ \rho(\phi^b), \pi(\phi^b) \}$ in such a way that one of them, say $\pi$, results neutral whilst the other, $\rho$, transforms as a shift under the above transformation. More precisely
\begin{equation}\label{eq:trans_weyl_2}
\rho'(x)  = \rho(x)  + M \sigma , \hspace{0.4cm} \pi'(x)  =\pi(x)
\end{equation}
The action, written in terms of such fields can only depend on derivatives of $\rho$. Therefore the potential $U$ and the field metric $\gamma_{ab}$ are functions of $\pi$ only. Furthermore, we require for $\gamma_{ab}$ to remain in diagonal form. By mean of the covariantly conserved current $J^\mu$ associated to the Weyl symmetry it is easy to find fields satisfying the conditions stated above.
%
Clearly, in terms of $\rho$ and $\pi$ such a current is given by
\begin{equation}\label{eq_current_1}
J^\mu = \frac{1}{\sqrt{-g}} \frac{\partial \mathscr L_E}{\partial \partial_\mu \phi^a} \frac{\delta \phi^a }{\sigma}= - M \gamma_{\rho \rho}(\pi) \partial^\mu \rho
\end{equation} 
where $\delta \phi^a$ is the infinitesimal variation of a field under \eqref{eq:trans_weyl_2} and $\gamma_{\rho \rho} (\psi)$ is the $\rho-\rho$ matrix element of the field metric. On the other hand, the same $J^\mu$ from \eqref{eq:L_fully_cov} can be computed in terms of the old fields $\phi$ and $f$. After some trivial steps one obtains
\begin{equation}\label{eq_current_2}
J^\mu = - M \left(\frac{f}{M}\right)^2 \partial^\mu \left[ \frac{\phi^2}{2 M} + 3M \left(\frac{M}{f}\right)^2\right]
\end{equation}
By direct comparison of the two results \eqref{eq_current_1}, \eqref{eq_current_2} and by imposing \eqref{eq:trans_weyl_2} it is easy to obtain a correct form for $\rho$, which we define as
\begin{equation}\label{eq:rho}
\rho \equiv \frac{M}{2}\log \left[ \frac{\phi^2}{2 M} + 3M \left(\frac{M}{f}\right)^2\right]
\end{equation}

Moreover, it is convenient to define $\pi$ as a function of $\phi f$. In order to obtain a canonical kinetic term for $\pi$ we decide to define this field as
\begin{equation}\label{eq:pi}
\pi \equiv M \sqrt{6}\; \text{arcsinh} \left[\frac{f \phi}{\sqrt{6}M^2} \right]
\end{equation}
 
Finally, from definitions \eqref{eq:rho}, \eqref{eq:pi} the action \eqref{conf_model_action} is rewritten as
\begin{equation}\label{inflat_action}
I_E = \int  \left[ \frac{M^2}{2 }R - \frac{1}{2}\partial_\mu \pi \partial^\mu \pi  - 3 \cosh^2 \left[ \pi / (\sqrt{6}M) \right]\partial_\mu \rho \partial^\mu \rho - V(\pi) \right] \sqrt{- g} \, \text{d}^4 x\,,
\end{equation}
where
\begin{equation}\label{poten_inflaton}
V(\pi) = \frac{9 \lambda M^4}{4 \xi^2}\left[ 1- 4\xi \sinh^2\left[ \pi / (\sqrt{6}M) \right] + 8 \xi^2\sinh^4\left[ \pi / (\sqrt{6}M)  \right]\right]
\end{equation}

Manifestly, the field $\rho$ represents the massless mode associated to the flat directions of the potential $U$, this is why it has no potential term.
Due to this fact, the inflationary analysis in section \ref{IV} will be greatly simplified.

\section{Global evolution}\label{III}

A viable inflationary model needs  a sufficiently long phase of quasi-exponential accelerated expansion for the scale factor $a(t)$. To see if such a phase is present, we study the equations of motion in a flat homogeneous and isotropic FLRW background with metric $\text{d}s^2 = -\text{d}t^2 + a(t)^2 \delta_{ij} \text{d}x^i \text{d}x^j$. 

First, we express our set of equations in terms of the e-fold number $N \equiv \log a(t)$ and we set $\alpha=\xi^{2}/\lambda$ in order to recover general relativity at late times (see sec. \ref{II}). Additionally, it is more convenient to work with dimensionless variables 
\begin{equation}
\label{dynamical_variables}
x \equiv \gamma \chi, \hspace{0.2cm} y \equiv \gamma \chi ', \hspace{0.2cm} z \equiv  \phi / M, \hspace{0.2cm} w \equiv \phi' / M, \hspace{0.2cm} h \equiv  H /M 
\end{equation}
with $'$ being a shorthand notation for $\text{d}/\text{d}N$ and $H \equiv \dot{a} /a$ being the Hubble function. By inserting these variables in the field equations \eqref{eq1_field}, \eqref{eq2_field}, and the $tt$-component of \eqref{conf_model_eq} we find the first-order coupled system of equations given by
\begin{align}\label{conf_dyn_system}
&x' - y = 0\,, \\
&y' + \left[\frac{h'}{h} +3\right] y +  \left[\frac{\lambda}{\xi}\frac{z^2}{h^2} + \frac{w^2}{3} \right]e^{-x} - \frac{2 \lambda}{3}  \frac{z^4}{h^2} e^{-2x} = 0\,, \\
& z' - w = 0\,, \\
& w' + \left[\frac{h'}{h} + 3 - y\right]w- \frac{3 \lambda}{\xi h^2}z + \frac{2 \lambda}{h^2} e^{-x}  z^3= 0\,,\\
&h^2 \left[1-\frac{y^2}{4} - \frac{w^2}{6}e^{-x}\right] = \frac{\Lambda}{3 M^2} +\frac{\lambda}{6}z^4 e^{-2x} - \frac{\lambda}{2\xi} z^2 e^{-x} \label{friedmann}\,.
\end{align}
Similarly to the case in the Jordan frame, this system admits two families of fixed points, namely solutions for the equation $(x', y', z', w' ) = 0 $, corresponding to
\begin{align}
(x, y, z, w) &= (x_1, 0, 0, 0)\,,   &h_1= \pm\frac{\sqrt{3 \lambda}}{2\xi}\,, \label{E_unst}\\
(x, y, z, w) &= \left(x_2, 0, \pm \sqrt{\frac{3}{2 \xi}} e^{\frac{x_2}{2}}, 0\right)\,,  &h_2 = \pm 	\frac{\sqrt{3 \lambda}}{2\sqrt{2}\xi}\,. \label{E_stable}
\end{align}
where $x_1$ and $x_2$ are arbitrary numbers. As it will be shown below, \eqref{E_unst} represents a saddle point, whereas \eqref{E_stable} is a stable point of the dynamical system. 

We now focus on the linearized solutions around those points. 
In this way we can assess the stability of the background equations of motion and verify the existence of inflationary solutions connecting a de Sitter Universe to a radiation-dominated Universe. 

\subsection{Unstable fixed point}

We perturbatively expand the variables $ x = x_1 + \delta x$  and $h = h_1 + \delta h$ around the fixed point \eqref{E_unst} as functions of $N$ only, keeping in mind that $\delta x$, $y$, $z$, $w\ll 1$ and $x_1$ is arbitrary. 
Retaining only linear terms in the equations and using the constant value of $h^2$ in the saddle point, we obtain $\delta h = 0$ identically. This is to be expected, since the Friedmann equation in the Einstein frame represents a constraint for the dynamical system. Additionally, the remaining equations for $\delta x$ and $z$ are solved by 
\begin{align}
\label{conf_unst_sol}
& \delta x(N) = c_1 e^{-3N} + c_2 \nonumber \\
& z(N) = e^{-\frac{3}{2}N} \left[c_3 e^{ -l N / 2}+ c_4 e^{  l N /2}\right]
\end{align}
where $l \equiv \sqrt{16\xi +9}$ and $c_i$ are constants of integration. Clearly $\delta x $ has a constant and stable behaviour whilst $ z = \phi / M$ drags the system away from equilibrium (if $c_4 $ is non vanishing). Once again, the solutions mimic the corresponding results found in the Jordan frame. 

\subsection{Stable fixed point}

To obtain the equations around the stable point, we set the new  variables $g$ and $q$ such that
\begin{equation}
\phi = e^g M \hspace{0.5cm} \phi' = q e^g M 
\end{equation}
and keep $x$ and $y$ as before so that 
\begin{align}
&(x,y,z,w)\rightarrow(2\bar{g} + \log (2\xi /3), 0, \bar{g}, 0 )\,,
\end{align}
whereas $\bar{g}$ has been defined in such a way that the arbitrary stable fixed point $x_2$ for $x$ is $x_2 = 2\bar{g} + \log (2\xi /3)$. 
We then expand our functions as $\gamma \chi = x_2 + \delta x$, $\phi = e^{\bar{g}}(1+\delta g) M $ and $h = h_2 + \delta h$. Plugging this parametrization in the above equations and keeping only linear terms, we find a system of two linear equations
\begin{align}
 &\delta x'' + 3 \delta x'+ 4 \left( \delta x - 2 \delta g \right) = 0 \label{conf_stable_system1}\\
 &\delta g'' + 3 \delta g' - 8 \xi \left( \delta x- 2 \delta g \right) = 0\label{conf_stable_system2}
\end{align}
Once again, equation \eqref{friedmann} yields $\delta h = 0$. Just by looking at \eqref{conf_stable_system1} and \eqref{conf_stable_system2} we see that they have a symmetrical form. Indeed, a particular solution of this system is given by $2 \xi \delta x = \delta g + c$, with $c$ constant. Actually, if it is required for the solution to approach the fixed point asymptotically for $N\rightarrow +\infty$, then it must be the case for $c$ to vanish. If we employ these results, then the system becomes disentangled and its solution can be obtained as a superposition of exponential functions
\begin{align}\label{conf_linear_st_sol}
&\delta g(N) =  e^{-3/2N} \left[C_1 \sin (K N /2) + C_2 \cos (K N /2)\right] \nonumber \\ 
& \delta x(N) = -\frac{ e^{-3/2N}}{2\xi} \left[C_1 \sin (K N /2) + C_2 \cos (K N /2)\right]
\end{align}
where $C_1$ and $C_2$ are integration constants and $K \equiv \sqrt{64 \xi + 7}$. A general solution of the linear system around this fixed point would require four different initial conditions. However, since we are working with a particular solution, only two of these  are necessary. This solution is manifestly stable and the same also applies to the most general solution, which shares similar oscillatory behaviour and damping factor.

\section{Inflationary phase}\label{IV}

The transition from a saddle point to a stable point allows for a phase of accelerated expansion $\ddot{a}/a> 0$ followed by a preheating stage of the Universe. Indeed, as we will see, with initial conditions close enough to the saddle point it is possible to obtain a long lasting inflationary trajectory (with $H$ almost constant) that  ends when the field $\phi$ starts to oscillate and  drives the system towards the stable equilibrium point. At this stage, preheating is triggered by the damped oscillations of $\phi$ and $\chi$ close to the stable fixed point. The damped oscillations of $\chi$ play the role of the damped oscillations of $H$ in the Jordan frame.

Thanks to the analysis of sec. \ref{II} we prefer to study inflation by mean of the action \eqref{inflat_action}. The behaviour of the fields $\rho$, $\pi$ around the fixed points can be easily obtained by using their definitions in terms of $\phi$ and $\chi$.

The nearly exponential inflationary expansion can be achieved if the acceleration condition is satisfied for a sufficiently long number of e-folds:
\begin{equation}
\epsilon_1 \equiv - \frac{H'}{H} = \frac{1}{2}\gamma_{ab} \; {\phi'}^a {\phi'}^b \ll 1
\end{equation}
where in the second equality we used the Friedman equation for the Hubble function and the equations of motion for the scalars.

Strictly speaking our model belongs to the class of two-field inflationary models \cite{wands2002observational}. However we will now argue that in the specific conditions necessary for inflation, the role of $\rho$ is completely negligible both at the level of background and perturbed equations. Therefore, effectively its degree of freedom can be neglected in our analysis.

For the background evolution (e.g. homogeneous and isotropic  limit) this can be seen as follows. The equation for $\rho$ can be obtained directly from the covariant conservation equation for the current \eqref{eq_current_1}. Indeed, the equation $\nabla_\mu J^\mu =0 $, when specified for $\rho = \rho(N)$, implies
\begin{equation}\label{rho_evolution}
\rho'' + \rho'\left[ 3+ \frac{H'}{H} + \pi' \frac{\partial}{\partial \pi}  \log \gamma_{\rho \rho}(\pi) \right] = 0
\end{equation}

As will be discussed later, if the system starts very close to the unstable fixed point \eqref{E_unst} then inflation can be successful. Therefore, it is easy to check that we must choose initial conditions of the type $\rho'(0)$, $\pi'(0) \ll 1$ which together with \eqref{rho_evolution} clearly imply $\rho'(N) \simeq 0 $. In other words, the linearized version of \eqref{rho_evolution} in the unstable region gives an exponentially decaying evolution for $\rho'(N)$ \footnote{Alternatively, it is possible to show that the slow-roll plus slow-turn approximation (defined for example in \cite{peterson2011testing}), at any order, implies $\rho(N) = const$. }.

The constant behaviour of $\rho(N)$, in principle, is not sufficient to claim that our model belongs to single-field inflation. In particular, we have to assess how the perturbation $\delta \rho(x)$ affects the relevant cosmological parameters in a quantitative way. From scale invariance, however, we expect this effect to be completely negligible. Indeed, under the hypothesis of scale invariance it has been proven in \cite{garcia2011higgs} that the contributions given by $\delta \rho(x)$ decrease exponentially fast with $N$. Since inflation is required to last at least $N \sim 50$ e-folds, then we are allowed to neglect completely the dynamics of $\rho$.  

In this regime, the relevant part of the action during inflation is
\begin{equation}\label{eff_action_inflation}
I_E =   \int \left[ \frac{M ^2}{2}R - \frac{1}{2}\partial_\mu \pi \partial^\mu \pi -  V(\pi)\right] \sqrt{-g} \, \text{d} ^4x\,,
\end{equation}
where  $V(\pi)$ has been defined in \eqref{poten_inflaton}. The expression \eqref{eff_action_inflation} has now the form of an inflationary action with a single canonically normalized scalar field, hence it can be studied with the usual methods in the slow-roll approximation (see e.g.  \cite{martin_encyclopaedia_2014,linde_inflationary_2008}).

\begin{figure*}[hbtp]
\centering
\includegraphics[scale=0.42, trim={0 0.1cm 0 0},clip]{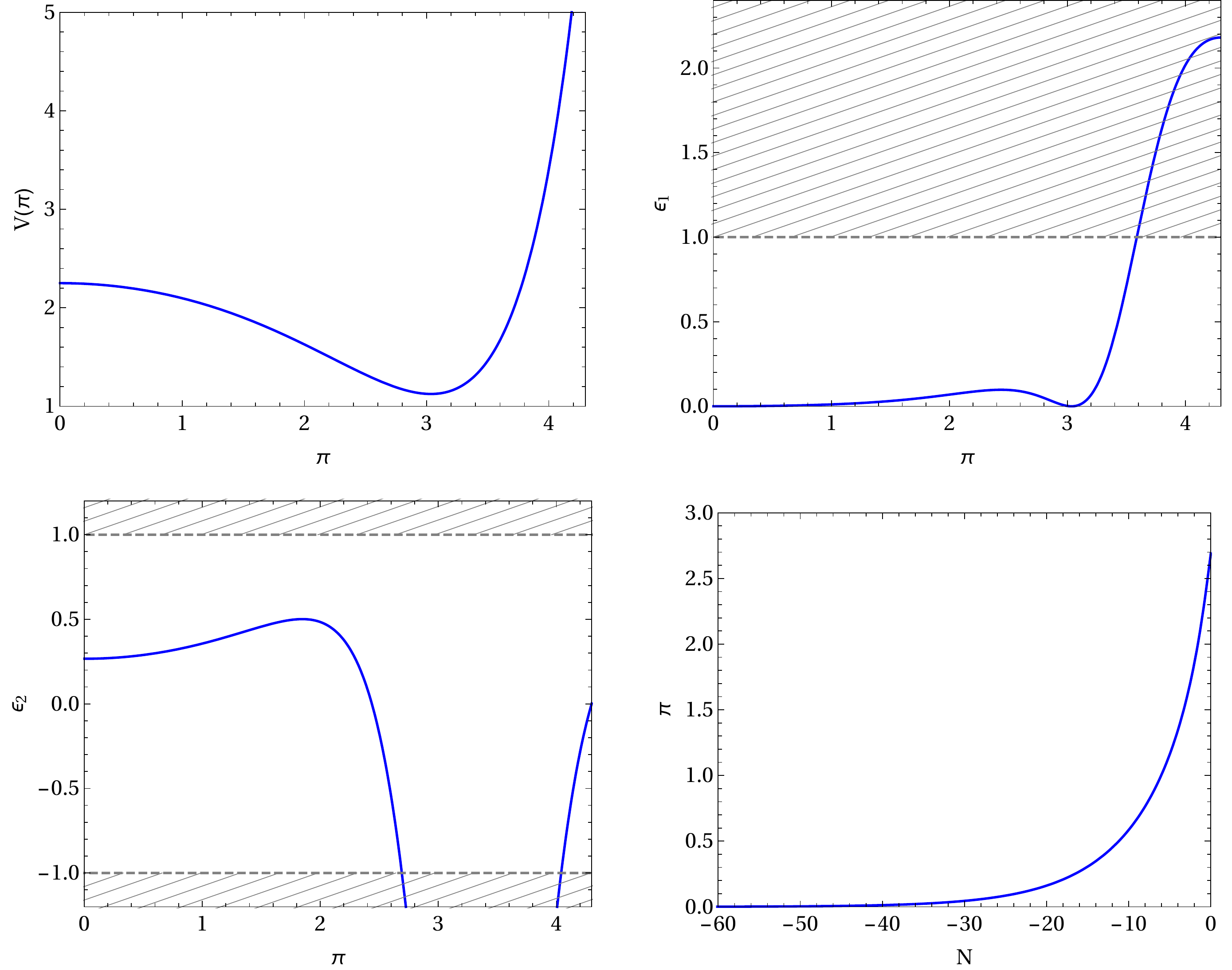}
\caption{Top left: potential for the inflaton $\pi(x)$ in the Einstein frame. Top right, and bottom left: slow-roll parameters $\epsilon_1$, and $\epsilon_2$ (equations \eqref{E1}, \eqref{E2}) as functions of the inflaton expectation value $\pi$. The shaded regions indicate where the slow-roll parameters become larger in modulus than 1. Bottom right: evolution for $\pi(N)$ in the slow-roll approximation from $N_* = -60$ to $N_{end} = 0$. In all the plots we choose $\xi =10^{-1}$, $\lambda = 10^{-2}$.}\label{fig1}
\end{figure*}

In particular, following \cite{liddle_formalizing_1994,schwarz_higher_2001}, we introduce the hierarchy of Hubble flow parameters defined by
\begin{equation}
\epsilon_{n+1} \equiv \frac{\text{d}}{\text{d}N} \log |\epsilon_n|, \hspace{0.4cm} \epsilon_0  \equiv \frac{H_{\text{in}}}{H}\,.
\end{equation} 
With this formalism the acceleration condition reads $\epsilon_1 < 1$ and inflation is conventionally considered over when $|\epsilon_i | = 1$ for either $i = 1$ or $i = 2$. Moreover, the slow-roll conditions require that all $|\epsilon_i |\ll 1$ \cite{martin_encyclopaedia_2014}. When such conditions are satisfied, the Friedman equation, together with the field equation for $\pi$, allow the $\epsilon_i$ to be expressed as functions of $V$ and its derivatives.
\begin{align}
\epsilon_1 &\simeq \frac{M^2}{2}\left[\frac{V_\pi}{ V} \right]^2  \,, \label{E1}
\\
\epsilon_2 &\simeq 2 M^2 \left( \left[   \frac{V_\pi}{V} \right]^2 - \frac{V_{\pi \pi}}{V} \right)\,,  \label{E2}
\\
\epsilon_3 \epsilon_2 &\simeq 2 M^4  \left(  \frac{V_{\pi \pi \pi}V_{ \pi}}{V^2}-3 \frac{V_{\pi \pi}}{V} \left[ \frac{V_{\pi }}{V}\right]^2 + 2 \left[ \frac{V_{\pi }}{V}\right]^4 \right) \,, \label{E3}
\end{align}
where the subscript indicates the derivative with respect to $\pi$. Since all these parameters are functions of logarithmic derivatives of the potential \eqref{poten_inflaton}, manifestly they do not depend on $\lambda$.

\begin{figure*}[hbtp]
\centering
\includegraphics[scale=0.42, trim={0 0.5cm 0 0.5cm},clip]{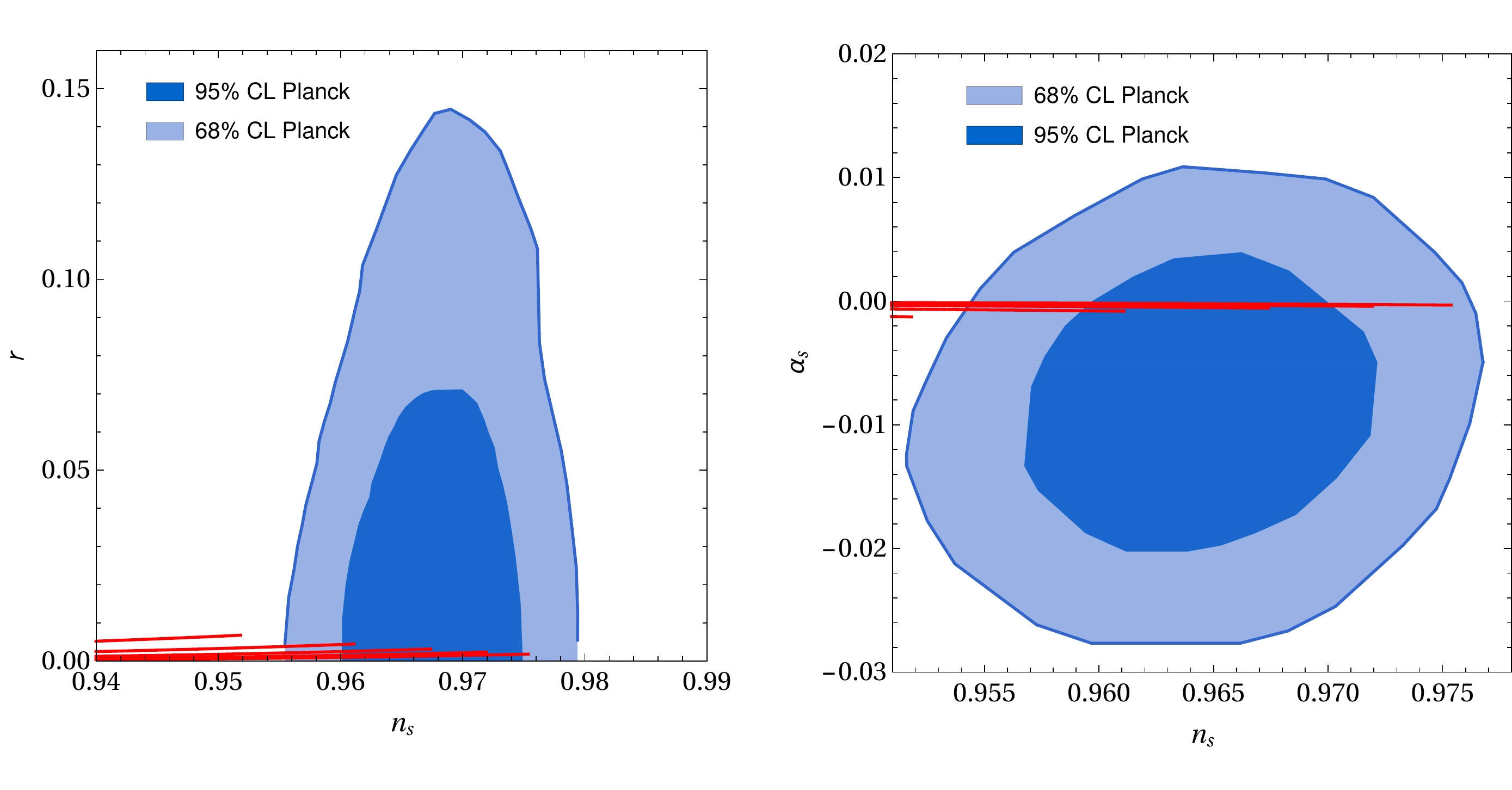}
\caption{Left: $(n_s, r)$  plane comparison between the scale invariant model (red), and the  $68\%$ and $95\%$ CL regions given by Planck \cite{planck_collaboration_planck_2015}. Right: $(n_s, \alpha_s)$ plane comparison for the same theoretical model and Planck results. In both plots each segment for the theoretical prediction is obtained, at a fixed $\Delta N_*$, by varying $10^{-5} < \xi < 5 \cdot 10^{-2}$. The different segments are obtained by setting $\Delta N_*= \{40, 50, 60, 70, 80\}$. On the left, curves with higher $r$ correspond to lower $\Delta N_*$, on the right the opposite.} \label{fig2}
\end{figure*}

\begin{figure*}[hbtp]
\centering
\includegraphics[scale=0.42, trim={0 1.7cm 0 1.6cm},clip]{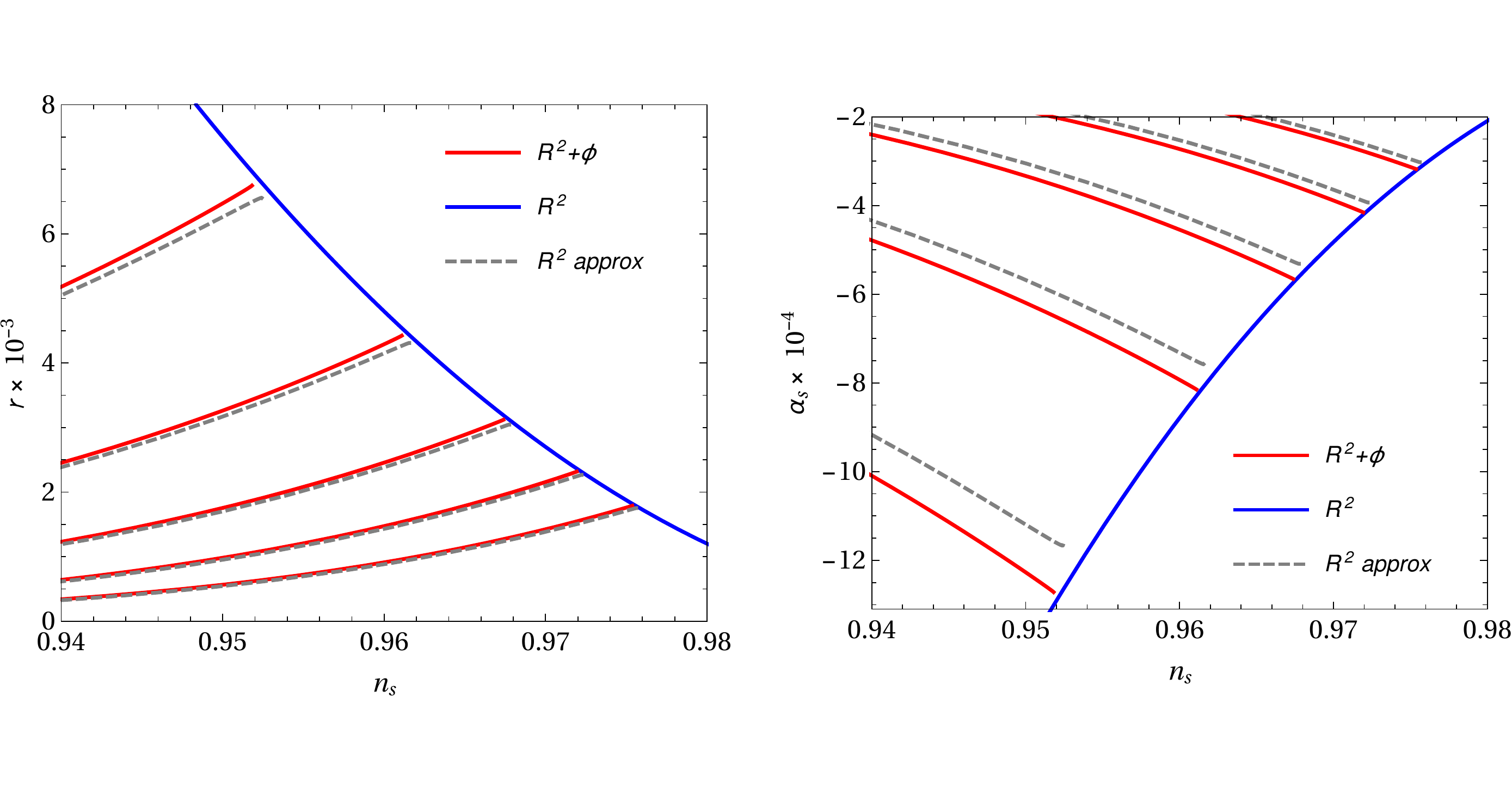}
\caption{ Comparison between the numerical results for the scale invariant model (red), the approximated results obtained with \eqref{pi_star} (dashed) and the Starobinsky model (blue) in the $(n_s, r)$ region (left) and $(n_s, \alpha_s)$ region (right). Left: different red curves ore obtained by fixing $\Delta N_*$ and by varying $10^{-5}< \xi < 5 \cdot10^{-2}$. The values taken by $\Delta N_*$ are, by starting from the top, $\Delta N_* = \{40, 50, 60, 70, 80\}$. The blue curve is obtained by mean of \eqref{staro_formula}. Right: same as before, where here the red curves start with $\Delta N_* = 40$ at the bottom. }\label{fig3}
\end{figure*}

\begin{figure*}[hbtp]
\centering
\includegraphics[scale=0.42, trim={0 0.5cm 0 0.5cm},clip]{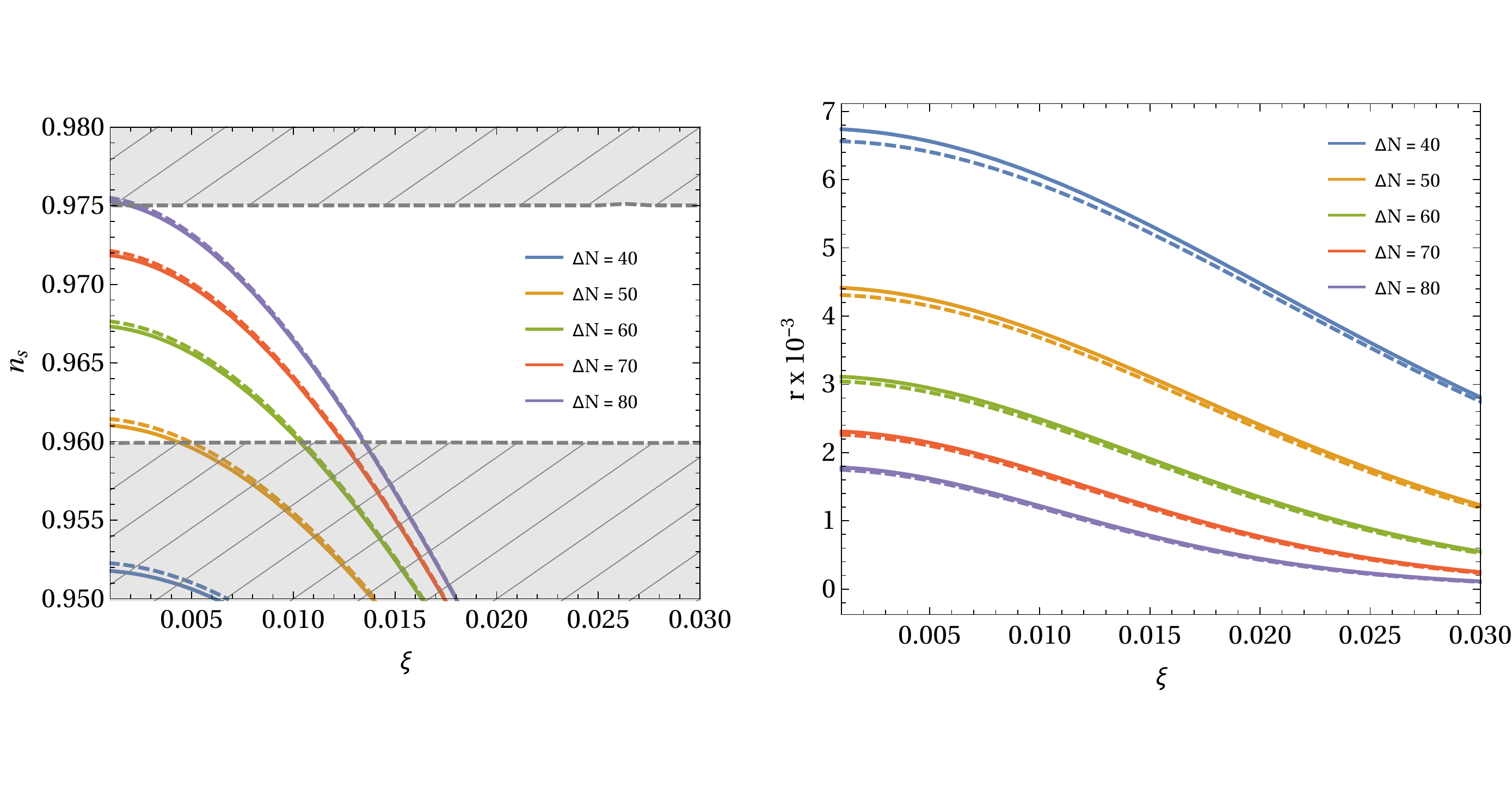}
\caption{Dependence of the spectral parameters on $\xi$ and $\Delta N_*$. Continuous lines are obtained numerically at second order in the slow-roll approximation, dashed lines are obtained through the approximation \eqref{pi_star}. Left: values of the spectral index $n_s$. The shaded region is excluded by the $68\%$ CL Planck results \cite{planck_collaboration_planck_2015}. Right: values of the tensor-to-scalar ratio $r$.} \label{fig4}
\end{figure*}

Our aim is to obtain second order results for the cosmological parameters, hence we focus only on the first three $\epsilon_i$'s. For example, the first parameter $\epsilon_1$ is given by
\begin{align}
\epsilon_1 &\simeq \frac{4 \xi ^2 \sinh ^2\left[\sqrt{\frac{2}{3}} \pi\right] \left[2 \xi -2 \xi  \cosh \left[\sqrt{\frac{2}{3}} \pi\right]+1\right]^2}{3 \left[8 \xi ^2 \sinh
   ^4\left[\frac{\pi}{\sqrt{6}}\right]-4 \xi  \sinh ^2\left[\frac{\pi}{\sqrt{6}}\right]+1\right]^2} \, \label{e1}
\end{align}
These parameters are shown in Fig [\ref{fig1}] as a function of $\pi$. When the slow-roll approximation holds, the evolution equation for the inflaton reduces to a first-order differential equation:
\begin{equation}\label{eq_slow-roll_inf}
\left( 1+  \frac{\epsilon_2}{6 -2 \epsilon_1}\right) \pi'(N) = - M \sqrt{2 \epsilon_1}
\end{equation} 

To allow for slow-roll we would like to find regions in the phase space of $\pi$, for which $\epsilon_1$, $\epsilon_2\ll 1$. This could be the case around the origin $\pi = 0$, where their approximate expressions are
\begin{equation}
\epsilon_1  \simeq \frac{8}{9}\xi^2 \pi^2 + \mathcal{O}(\pi^4 ), \hspace{0.5cm}
\epsilon_2  \simeq \frac{8}{3}\xi +\epsilon_1 +  \mathcal{O}(\pi^4 )\\
\end{equation} 
We see that $\epsilon_2$ has a finite value for $\pi$ exactly zero. Therefore, only in the case $\xi< 3/8$ the slow approximation can be sensible around the origin.
Finally, from Fig [\ref{fig1}] it is clear that there are no other regions where slow-roll is possible. In other words, the potential $V$ for large values of $\pi$ becomes too steep to allow for slow-roll. We conclude that initial conditions close to the unstable fixed point (where $\pi = 0$) are able to trigger slow-roll inflation.

In all these situations, $|\epsilon_2| = 1$ happens before $\epsilon_1 = 1$. Hence, the former condition fixes the value of the inflaton at the time $N_{end}$ when inflation ends $\pi(N_{end}) = \pi_{end}$. Generically, $\pi_{end}$ is a complicated function of the parameter $\xi$. However, from the previous discussion it is reasonable to consider the limit $\xi \ll 1$ were we obtain an analytical result:
\begin{equation}\label{inflaton_end}
\pi_{end} \simeq \sqrt{6}\; \text{arcsinh} \left[ \frac{\sqrt{3+8 \xi}}{4 \sqrt{\xi}} \right]
\end{equation}
In order to obtain analytical expressions we will work with the assumption $\xi \ll 1$ in what will follow. In our numerical results for the cosmological parameters, however, this approximation is not implemented. 

In the regime $\xi \ll 1$ the evolution equation for $\pi$ can be easily solved. Specifically, if we define $\Delta N_*$ as the difference between $N_{end}$ and the horizon crossing time $N_*$, meaning that $\Delta N_* \equiv N_{end} - N_*$ then \eqref{eq_slow-roll_inf} can be integrated to find $\pi_* \equiv \pi(N_*)$. We find
\begin{align}
\Delta N_* &= -\frac{1}{M}\int_{\pi_*}^{\pi_{end}} \left[1 +\frac{\epsilon_2}{6 - 2 \epsilon_1}\right]\frac{ \text{d} \pi}{\sqrt{2 \epsilon_1}} \,  \nonumber \\ 
&\simeq -\frac{3}{4 \xi M} \log \left[ \frac{\tanh (\pi_* / \sqrt{6})}{\tanh (\pi_{end} / \sqrt{6})}\right] +  \mathcal{O}(\xi^0)
\end{align}
Then, thanks to equation \eqref{inflaton_end} we get the value of $\pi$ at horizon crossing
\begin{equation}\label{pi_star}
\pi_* \simeq \sqrt{6} M  \, \text{arctanh}\left[\sqrt{\frac{8 \xi +3}{24 \xi +3}} e^{-\frac{4 \xi  \Delta N_*}{3}}\right]
\end{equation}

At this point we are ready to characterize the scalar and tensor power spectra of cosmological perturbations, indicated respectively as $\mathscr{P}_{\mathcal{R}}$ and $\mathscr{P}_t$. These quantities need to be evaluated at the conformal time $\eta_*$ at which the pivot comoving wave number $k_*$ crosses the Hubble horizon. Of particular interest are the spectral indexes \cite{mukhanov_theory_1992,martin2003inflation}
\begin{equation}
n_s \equiv 1+\left. \frac{\text d \log \mathscr{P}_{\mathcal{R}}}{\text d \log k} \right|_{k_*}, \hspace{0.4cm} n_t \equiv \left. \frac{\text d \log \mathscr{P}_{t}}{\text d \log k} \right|_{k_*},
\end{equation}
their runnings, and the tensor-to-scalar ratio
\begin{equation}
\alpha_s \equiv \left. \frac{\text d ^2\log \mathscr{P}_{\mathcal{R}}}{(\text d \log k)^2} \right|_{k_*}, \hspace{0.2cm} \alpha_t \equiv \left. \frac{\text d^2 \log \mathscr{P}_{t}}{(\text d \log k)^2} \right|_{k_*}, \hspace{0.2cm}r \equiv \left. \frac{\mathscr{P}_t}{\mathscr{P}_\mathcal R}\right|_{k_*}.
\end{equation}
All these coefficients need to be expressed in terms of the slow roll parameters $\epsilon_i$ evaluated at $\pi_*$. By doing so we obtain the predictions of the scale invariant model at second order (see \cite{martin_k-inflationary_2013, beltran_jimenez_exact_2013, casadio_higher_2005,  choe_second_2004, martin_wkb_2003, casadio_improved_2005-1, stewart_density_2001})

\begin{align}
n_s &= 1 -2 \epsilon_1 - \epsilon_2 - 2 \epsilon_1^2 - C \epsilon_2 \epsilon_3 -(2 C+3) \epsilon_1 \epsilon_2 \label{ns}
\\
n_t &= -2 \epsilon_1 - 2 \epsilon_1^2 -2 (C+1)\epsilon_1 \epsilon_2 \label{nt}
\\
\alpha_s & = -2 \epsilon_1 \epsilon_2 - \epsilon_2 \epsilon_3 \label{alphas}
\\
\alpha_t & =-2 \epsilon_1 \epsilon_2 \label{alphat}
\\
r & = 16 \epsilon_1 \left[ 1+ C \epsilon_2 + \left(C - \frac{\pi^2}{2} + 5\right)\epsilon_1 \epsilon_2 + \left( \frac{C^2}{2} - \frac{\pi^2}{8} + 1 \right) \epsilon_2^2 + \right. \nonumber \\ &\left. +  \left( \frac{C^2}{2}- \frac{\pi^2}{24}\right)\epsilon_2 \epsilon_3 \right]
\label{r}
\end{align}
where $C \equiv \gamma_E + \log 2 - 2 \approx -0.7296$ and $\gamma_E$ is the Euler constant. The slow-roll parameters in these expressions can be obtained with a very good accuracy by using the expression \eqref{pi_star} for $\pi_*$. At lowest order they are approximately given by
\begin{align}
\epsilon_1 &\simeq \frac{16 \xi ^2 (8 \xi +1) (8 \xi +3) e^{-\frac{8 \xi  \Delta N_*}{3}}}{\left[(8 \xi +3) e^{-\frac{8 \xi  \Delta N_*}{3}}-3 (8 \xi +1)\right]^2} \label{ee1}
\\
\epsilon_2 &\simeq -\frac{8 \xi  \left[24 \xi +8 \xi  e^{-\frac{8 \xi  \Delta N_*}{3}}+3 e^{-\frac{8 \xi  \Delta N_*}{3}}+3\right]}{3 \left[-24 \xi +8 \xi  e^{-\frac{8 \xi  \Delta N_*}{3}}+3 e^{-\frac{8
   \xi   \Delta N_*}{3}}-3\right]} \label{ee2}
\\
\epsilon_3 & \simeq -\frac{16 \xi  (8 \xi +1) (8 \xi +3) e^{-\frac{8 \xi \Delta N_*}{3}}}{(8 \xi +3)^2 e^{-\frac{16 \xi  \Delta N_*}{3}}-9 (8 \xi +1)^2} \label{ee3}
\end{align}

The numerical and analytical results for the spectral indexes $n_s$, $\alpha_s$, and $r$ are shown in Fig [\ref{fig2}-\ref{fig4}] together with the corresponding experimental bounds given by Planck \cite{planck_collaboration_planck_2015}.

One of the most solid inflationary model at the moment is the one introduced by Starobisnky. In particular, its prediction for the spectral index $n_s$ and for the tensor-to-scalar ratio $r$ is parametrized by 
\begin{equation}\label{staro_formula}
n_s \simeq 1 - \sqrt{r/3}
\end{equation} 
Our results are expected to converge to those obtained by Starobinsky for $\xi \rightarrow 0 $. In this limit, the original scalar field $\phi$ results minimally coupled to gravity and does not influence inflation. This indeed is the case, as it is clearly visible in Fig [\ref{fig3}] in the case of $n_s$. Similar conclusions are also drawn when one specializes to $\alpha_s$ instead.

More interesting features appear for $\xi \neq 0 $. In this case our model, compared to the Starobisnky one, is able to span a larger region in the parameter spaces $(n_s, r)$ and $(n_s, \alpha_s)$, which are strongly constrained by current observations. Therefore, from the experimental results we are able to put bounds on the parameters $\xi$ and $\Delta N_*$. Interestingly enough, Fig [\ref{fig4}] shows that our model is within the experimental constraints $0.96 \lesssim n_s \lesssim 0.975$ for $\xi\lesssim 1.5 \cdot 10^{-2} $ and $50 \lesssim \Delta N_* \lesssim 80$. For these values of $\xi$, generically $n_s$ is smaller than in Starobinsky's inflation.

Finally, we can convey that \eqref{model_action} is a perfectly viable inflationary action. Moreover, this analysis suggests that a nearly scale invariant spectrum of perturbations cannot be obtained when large non-minimal couplings are present in quadratic theories of gravity.

\section{Preheating} \label{V}

After inflation has ended, the system approaches a stable configuration through damped oscillations, described by eqs.\  \eqref{conf_linear_st_sol}, which allow an energy transfer from the two scalar fields to the Standard Model fields, opening the way to a preheating phase of the Universe (\cite{kofman_towards_1997}, \cite{doi:10.1146/annurev.nucl.012809.104511}). Without an effective amplification of the Standard Model fields (and the consequent realization of a thermal state), our model would be doomed. Indeed, the Hubble function converges to the constant value \eqref{E_stable}, which is incompatible with a subsequent radiation era. However, an efficient mechanism of particle production can fill the Universe with massless radiation that takes over the matter content of the Universe and drive it towards the standard radiation-dominated epoch. In order to check whether such a mechanism can arise, we study the dynamics of a new scalar field $\psi$, minimally coupled to the metric, as representative of a Standard Model field. We further postulate a scale invariant interaction between $\psi$ and $\phi$, with dimensionless coupling constant $g^2$ (here we do not exclude the possibility of having $g^2<0$). With these hypothesis, the effective action  during preheating is
\begin{equation}
I_E = -\int \left[ \frac{1}{2}  e^{-\gamma \chi} \partial_\mu \psi \partial^\mu \psi+ \frac{g^2}{2} e^{-2\gamma \chi} \psi^2 \phi^2  \right] \, \sqrt{-g} \, \text{d}^4x\,,
\end{equation}
where the exponential factors $e^{-\gamma \chi}$, $ e^{-2\gamma \chi}$ appear because we are working in the Einstein frame. For convenience, we define a new field $f$ (whose Fourier transform is $f_k(t)$) in such a way that $\psi_k(t) = a^{-3/2} e^{\gamma \chi /2} f_k(t)$. Finally, we use the e-fold number $N$ as time together with the variables defined in \eqref{dynamical_variables}. As a result we find the equation
\begin{equation}\label{preheating_eq}
f_k'' + \frac{h'}{h} f'_k + \Omega_k^2 f_k = 0\,,
\end{equation}
where the time-dependent frequency $\Omega_k$ is 
\begin{equation}\label{Om}
 \Omega^2_k \equiv p_k^2 + \frac{g^2 z^2}{h^2}e^{-x} +  \frac{x'}{2}\left[ \frac{h'}{h} + 3\right] - \frac{x'^2}{4} + \frac{x''}{2}  - \frac{3}{2} \frac{h'}{h} - \frac{9}{4}  \,,
\end{equation}
and $p_k \equiv k  / (a h M)$. This differential equation needs to be solved once the homogeneous solutions for $x$, $z$, and $h$ around the stable fixed point \eqref{conf_linear_st_sol} are provided. However, analytical solutions can be obtained only in certain approximations. In this respect, we first note that $h'/h$ is very small close to the stable configuration (this is also verified in numerical calculations) thus can be safely neglected in the following discussion. Furthermore, since preheating is supposed to last for very few e-folds,  the damping factor in \eqref{conf_linear_st_sol} can be safely neglected. With these considerations, $x$ and $z$ can be written as 
\begin{equation}
x = x_0 + \bar{x} \sin \left(N K /2 \right)\,, \hspace{0.4cm}
z = z_0 + \bar{z} \sin \left(N K /2 \right)\,.
\end{equation}

The field amplification of $f_k$ is expected to happen in the so-called \emph{non-adiabatic limit}, where the frequency $\Omega_k$ varies quickly with respect to $N$. Indeed, in this regime the equation under investigation can present an exponential growth for certain $k$'s. In our model  this process is allowed in two different ways that we call $\chi$-amplification and $\phi$-amplification.

\subsection{$\chi$-amplification}

Lets assume that, after inflation, the amplitude of the $x$ oscillations is much larger that the one of $z$ (i.e. $\bar{x} \gg \bar{z} $). This approximation is forbidden by solution \eqref{conf_linear_st_sol} since in general $\xi > 1$ but, as already argued, such expression represents just a particular solution of the equations, and in general additional modes with different amplitudes appear.
Further, we consider $\bar{x}$ to be of order one and note that $K \gg 1$, so the relevant terms in the frequency \eqref{Om} are
\begin{equation}\label{chi_freq}
\Omega_k^2 \simeq p_k^2  - K^2 d \left[\sin(NK/2)  + d - d^2 \sin^2(N K /2)\right]\,,
\end{equation}
where $d \equiv \bar{x}/4$. For small values of $p_k$ the adiabatic limit can be broken when the last three terms in \eqref{chi_freq} vanish. This happens when $N$ is such that 
\begin{equation}
\frac{\bar N K }{2} = \arcsin\left[ \frac{1\pm \sqrt{1+ 4d^2}}{2d}\right] + 2\pi n \hspace{0.5cm} n \in \mathbb{Z}
\end{equation}
Our choice is to consider the $+$ sign solution for $n=0$. Equation \eqref{chi_freq} can be expanded around this $\bar{N}$ at first order, considering $p_k  \ll 1$ and constant. This yields the equation
\begin{equation}\label{eq_x_dominant}
f_k''(\tau) + \left(p_k^2 - \tau \mathscr A^2\right) f_k(\tau) = 0
\end{equation}
We have introduced the variable $\tau \equiv (N - \bar N)$ and the constant
\begin{equation}
\mathscr A^2 \equiv \frac{K^3}{2 \sqrt{2}}\sqrt{(1+4d^2)^{\frac{3}{2}} - (1+4d^2)}
\end{equation}

The general solution of \eqref{eq_x_dominant} is expressed in terms of Airy functions of first and second kind \cite{abramowitz1964handbook}
\begin{equation}
f_k(\tau) = C_1 \text{Ai}\left( s(\tau) \right) + C_2 \text{Bi}\left( s(\tau) \right) 
\end{equation}
with $s(\tau) \equiv \left( \mathscr A^2 \tau - p_k^2 \right)\mathscr A^{-4/3}$.
The $\text{Ai}$ function is exponentially suppressed for large and positive values of $s$, whilst it oscillates for negative values of the argument. On the contrary, $\text{Bi}$ has an exponential growth for positive arguments.
Thanks to this unstable behaviour, in the asymptotic limit for late times our solution has the approximate from 
\begin{equation}
f_k(\tau) \simeq  \frac{C_2}{\sqrt{\pi \sqrt{s(\tau)}}}\exp\left[ \frac{2}{3}s^{3/2}(\tau)\right]
\end{equation}
which results a good approximation for $s(\tau) \gtrsim 2 $. 
For reasonable values of $\bar x  \sim 1$ we obtain that $s(\tau) $ becomes of order one when $\tau \sim (8 \sqrt{\xi})^{-1}$, which in general is a really small value. Therefore, the power series in $\tau$ utilized in \eqref{eq_x_dominant} is still valid even when the asymptotic limit in $s(\tau)$ is taken.

The suppressed amplification for modes far within the horizon ($k\gg a H$) is explained by our solution: for large $p_k$ the argument of the Airy functions becomes negative, thus providing a stable oscillatory behaviour.

\subsection{$\phi$-amplification}
The remaining limit to analyze is the case $\bar{z} \gg \bar x$, where the oscillations of $\phi$ drive preheating. In this case the adiabatic approximation is violated for $\phi \rightarrow 0 $. This situation is possible only if $\bar z$ is at least as large as $z_0$ so we study the limit $\bar{z}\gg z_0$ and  we consider a small coupling constant so that $g z_{0} \ll 1$.
The field $\phi$ vanishes at $N = 0 $, or at a point $N = \bar{N}$ if the sine function has a phase. We then expand equation \eqref{preheating_eq} around $\bar N$, obtaining 
\begin{equation}\label{eq_z_dominant}
f_k''(\tau) + \left( \upsilon_k^2  + \tau^2 \zeta^2 \right) f_k(\tau) = 0\,,
\end{equation}
where we define again $\tau \equiv N- \bar N$, $\upsilon_k^2 \equiv p_k^2 + \tilde g^2 z_0^2 h^{-2} $, $\zeta \equiv \tilde g \bar z K / (2 h)$ and $\tilde g \equiv g e^{-x_0 / 2}$.
Actually, equation \eqref{eq_z_dominant} has the same form of the Schr\"odinger equation for the scattering of a particle through a parabolic potential (with opposite sign). Its general analytic solution is given by a combination of parabolic cylinder functions \cite{abramowitz1964handbook}.

This equation has been thoroughly studied in the context of preheating in the past and it is known to lead to a conspicuous particles production. Indeed for $g^2>0$, the periodic scattering through the potential barrier produces amplification as a superposition of many small increments. From the analytical solution one can obtain the ratio between the number densities $n_\psi$ and $n_\phi$ of the reheated field $\psi$ and of the scalar field $\phi$ respectively (see \cite{kofman_towards_1997}, \cite{mukhanov2005physical} for details). After multiple $\phi$ oscillations we obtain $n_\psi / n_\phi \sim 3^{NK/2\pi} $.
This exponential behaviour suggests that few e-folds are enough to transfer most of the inflaton energy into reheated fields. Our analysis, however,  breaks down after few oscillations when the energy densities of $\phi$ and $\psi$ become of the same order. From that point on back-reaction effects on the $\phi$ field are to be expected, thus solution \eqref{conf_linear_st_sol} no longer holds.

Interestingly, the limit $\bar z \gg \bar x$ also grants the possibility for a tachyonic preheating scenario with an even broader instability. Indeed, for $g^2<0$ the analytical solution of \eqref{eq_z_dominant} has a exponential asymptotic behaviour for late times
\begin{equation}
f_k(\tau) \simeq C_k \tau^{- \frac{1}{2}-\frac{\upsilon_k^2}{2 |\zeta|}} \exp\left[\frac{|\zeta| \tau^2}{2} \right] \,,
\end{equation}
with $C_k$ being a coefficient depending on $k$. The field amplitude grows effectively only after $\tau \sim \zeta^{-1/2}$ e-folds. We require for this value to be small in order to not spoil the validity of the series expansion in \eqref{eq_z_dominant}. For typical values of the parameters $\lambda < 10^{-1}$, $\xi > 1$, and $\bar{z} \sim x_0 \simeq 1$ we have $ \zeta^{-1/2} \ll 1$ for a vast range of $g$ values. After this amount of time, however, modes with large $k$ are not amplified as effectively as smaller modes. For instance, from the explicit form of $C_k$ it can be shown that the amplitude of larger modes at $\bar \tau = \zeta^{-1/2}$ is suppressed exponentially compared to a smaller mode $q$
\begin{equation}
\left| \frac{f_k(\bar \tau)}{f_q(\bar \tau)} \right| \sim 2^{-\frac{p_k^2 - p_q^2}{4 \zeta}}\,.
\end{equation}
Similar results also apply for the comoving number density and energy density of the amplified field.

\section{Conclusions}\label{VI}

In this paper we investigated, in the Einstein frame, the inflationary properties of a minimal model for quadratic gravity plus a non-minimally coupled scalar field with spontaneously broken scale symmetry.

A sufficiently long inflationary epoch is achieved as the dynamical system flows from an unstable to a stable configuration where the underlying scale symmetry is broken. As usual, in the Einstein frame, the scalar degree of freedom associated with the quadratic Ricci scalar term  in the original action in the Jordan frame, plays the role of the inflaton.

We first computed the spectral indexes as well as their runnings and we found that they are fully consistent with the latest observational constraints for mild values of $\xi$. Moreover, as an upshot of scale invariance, they turn out to be independent on the  parameters $\lambda$ and $M$.
In the minimally-coupled limit, our model reduced to Starobinsky's inflation. When this is not the case, we have shown that also non-minimally coupled scalars can be compatible with current observations.


In the second part of the paper we studied possible reheating mechanisms. We found that, when the system approaches a stable configuration after inflation, several preheating channels allow for a quick energy transfer to the Standard Models fields. Indeed, an exponential amplification can be triggered by both $\chi$ and $\phi$ with a similar efficiency, even when tachyonic couplings are present. We conclude that the reheating mechanism is satisfactory in the Einstein frame and that our model is viable. 

There are however several questions that naturally arise concerning this model. For instance, the inflationary phase has been studied in a classical limit, where thanks to the conservation of the current $J^\mu$ one scalar field becomes a spectator. Nonetheless, technically speaking the symmetry associated to this current is generically anomalous and it would be interesting to study how quantum corrections modify the inflationary behaviour.


Another interesting possibility is the generalization of the minimal scale-invariant action \eqref{model_action} to include other scale-invariant operators or a conformally invariant electromagnetic field to see if some kind of warm inflation \cite{berera1995warm} is also possible. We leave these questions for future work.

\begin{acknowledgements}
We thank L. Vanzo and S.  Zerbini for valuable discussions. G. T. wishes to thank the Department of Physics of the University of Trento, where part of this work was done.
\end{acknowledgements}
\newpage

\bibliographystyle{ieeetr}
\bibliography{paper_s}

\end{document}